# Enhancement of low-spatial-frequency components by a new phase-contrast STEM using a probe formed with an amplitude Fresnel zone plate

Masato Tomita [a,*] Yukinori Nagatani [a,*] Kazuyoshi Murata [a] and Atsushi Momose [b]

[a] National Institute for Physiological Sciences, Okazaki, Aichi, Japan,

[b] Institute of Multidisciplinary Research for Advanced Materials, Tohoku University, Miyagi, Japan

## Abstract

Electron microscopy is a powerful tool for visualizing the shapes of sub-nanometer objects. However, contrast is not in proportional to density distribution, and therefore achieving a quantitative understanding of specimens is not straightforward, especially for low-contrast subjects such as biological specimens. To overcome this problem, we have developed a new phase-contrast scanning transmission electron microscope (STEM) in which a probe beam formed with an amplitude Fresnel zone plate (FZP) and the resulting interference patterns produced by the zeroth and first order diffracted waves generated by the FZP are detected. We name it FZP-PC-STEM hereinafter. The amplitude FZP was manufactured by using focused ion beam (FIB) equipment, and the diffraction data were collected by using diffraction imaging technique. The validity of our proposed optical model was confirmed by comparing experimental and simulated images. Observations of carbon nanotube (CNT) bundles by this method showed that the contrast of low-spatial-frequency components in the CNT image was enhanced, unlike the case in conventional bright-field STEM. This method does not, in principle, require the post-image processing used in the diffraction imaging method, and it can be easily introduced into a conventional STEM system without major modifications. The stability and robustness of the method toward intense electron irradiation during long-time operation were also confirmed. We expect that the FZP-PC-STEM will be widely applicable to quantitative observations of radiation-sensitive light-element specimens, with simple and easy operation.

## 1.   Introduction

In one of the latest technologies for cryo-electron microscopy, a transmission electron microscope (TEM) equipped with a Volta phase plate (VPP) [1] is widely used to increase the image contrast. The VPP is especially effective in imaging radiation-sensitive or low-contrast specimens, such as small proteins or membrane proteins solubilized by detergents [2]. This method is an extension of the Zernike phase-contrast TEM (ZPC-TEM), which uses a Zernike phase plate (ZPP) [3]. The ZPP was developed to increase the image contrast of biological specimens, which are considered to be weak-phase objects. However, because the ZPP is placed in the back focal plane of the TEM objective lens and is thus irradiated by a high-density electron beam, unstable performance occurs as a result of charging and contamination. The VPP method was proposed to alleviate this problem, by using electric charging to realize the function of the ZPP. The VPP method is commercially available and has been used to obtain a series of results in recent years [2]. However, it is also known that the electric charging, and therefore the phase shift, changes gradually as a result of electron-beam irradiation during data collection [3]. Consequently, the VPP is automatically shifted by PC-control after a certain period of operation. Electron holography, another technique for phase observation, requires a vacuum region (a vacant area) to permit a reference wave to pass through a specimen. The method is efficient for observations of specific samples, for example, magnetic field distributions in fine particles or electric-field observations of transistors [4]. However, for biological specimens in cryo-EM, where thousands of biomacromolecules are embedded in a uniform thin film of ice on a holey grid, it is difficult to form a reference wave for every embedded biomacromolecule.

---

[*] Corresponding authors.
E-mail addresses: Masato Tomita (tomitama@nips.ac.jp) and Yukinori Nagatani (nagatani@nips.ac.jp) at National Institute for Physiological Sciences, 38 Nishigonaka Myodaiji, Okazaki, Aichi, Japan.

Based on the reciprocity between TEM and STEM, several designs of phase-contrast STEMs (PC-STEMs) have also been proposed [5]. Among these, matched-illumination detector interferometry (MIDI)-STEM [6] has been the most systematically studied. These studies showed that a low-spatial-frequency component of the specimen can be emphasized by placing a patterned phase plate at the probe-forming (condenser) aperture and by using a virtual detector that matches the illumination pattern. Yang et al. [7] reported that MIDI-STEM combined with ptychography permitted the simultaneous observation of light and heavy elements by combining phase information extracted by the MIDI method with $Z$-contrast imaging.

The method that we propose in this report is similar to MIDI-STEM in terms of the device configuration. However, we use an amplitude FZP as a probe-modulation device at the condenser aperture position. An amplitude FZP is relatively easy to fabricate and is resistant to degradation, whereas MIDI-STEM uses a phase FZP of a SiN film-based phase-modulation type that is not robust to an electron beam. Our data-processing procedure also differs from that for MIDI-STEM. Here, we describe the principle of image generation and we report a proof-of-principle experiment in comparison with computer simulations to demonstrate the validity of our proposed optical model. The images produced by our method (which we have named FZP-PC-STEM) for carbon nanotube (CNT) bundles showed significantly enhanced low-spatial frequency components than those produced by conventional bright-field(BF)-STEM.

## 2. Materials and Methods

### 2.1 Fabrication of the Fresnel Zone Plate

FZPs for X-rays are normally produced by a lithographic process. However, because FZPs for electron beams need to have finer structures than those for X-rays, we used the focused-ion-beam (FIB) method to manufacture an amplitude FZP in a 30 nm-thick SiN film mounted in a 0.1 x 0.1 mm window opened in a 3 mm-diameter Si disk (TX301X; Norcada Inc., Edmonton, AB). Before FZP fabrication, gold was deposited by radio-frequency (RF) sputtering (SVC-700LRF; Sanyu Electron Co. Ltd., Tokyo) on the top and bottom surfaces of the SiN film at thicknesses of 30 and 15 nm, respectively, to suppress charging. The deposition was performed at a rate of 20 nm/min with an RF power of 50 W in an argon atmosphere at 0.2 Pa pressure. TEM observation showed that the average diameter of the gold grains was about 20 nm.

Initially, we used a He-ion microscope (ORION NanoFab, Carl Zeiss Microscopy GmbH, Jena) for FIB processing [8], because this permits processing of SiN films at the nanometer scale. However, it took about ten hours to fabricate a 20 μm-diameter FZP with this equipment. Eventually, the formed pattern was deformed or severely damaged due to specimen drift and ion-beam instability during fabrication. We therefore selected a Ne ion source in the same microscope, because this permitted fabrication with a much shorter operating time. The ion beam was scanned on the SiN film by using the nanopatterning and visualization engine (NPVE) in the NanoFab system. The FZP pattern was designed with a 2048 x 2048 pixels bitmap. The scan size on the film was 18.9 x 18.9 μm. One scan of the pattern was performed with a beam dwell time of 10 μs (55 s/scan), and the scan was repeated about 20 times to produce the target pattern. Because it takes a larger dose to create the outer fine slits than the inner slit, the fabrication was stopped manually after we had confirmed that the holes were completely formed in the outermost slits.

The secondary electron image of the fabricated FZP is shown in Fig. 1. The focal length ($f$) of the FZP was designed to be 500 mm for 200 kV accelerated electrons. The diameter of the FZP was 18.9 μm and the outermost slit width was 64.5 nm. The innermost ring was connected by 12 bridges to the second innermost ring. To avoid losing modulation in specific directions, the bridges were rotated every three rings. Bridges were also added to narrow slits on the same radial lines to prevent the deformations that occurred in the outer thin slits. The aperture ratio of the FZP was designed to be 30% and the measured value was about 27% in TEM observations. Lens functions, such as convergence and divergence of the diffracted waves, were experimentally confirmed by changing the focus of the intermediate lens of the STEM. Calculations based on the FZP model were also consistent with the behavior of the diffraction patterns as shown in Appendix B.

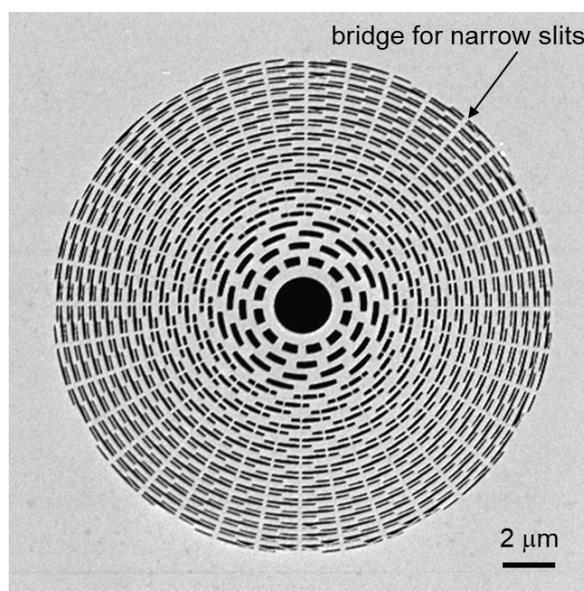

Fig. 1 Secondary electron image of the FZP fabricated on a Au/SiN film by using a Ne ion source with a He ion microscope

## 2.2 The sample and the imaging system of the FZP-PC-STEM

Multilayer carbon nanotubes (CNTs; Sigma-Aldrich, St Louis, MO) with diameters of 2.3–3.0 nm and lengths of 2–6 mm were observed as test samples in the FZP-PC-STEM. The CNTs were subjected to ultrasonic dispersion for ten minutes in pure water and the resulting sample was collected on a holey grid.

In this experiment, CNTs on an amorphous supporting film (holey grid) were selected because this system is comparatively stable under electron-beam irradiation.

The FZP was mounted on a condenser aperture holder and inserted into the column of the STEM (JEM2010F equipped with a 200 kV field-emission gun; JEOL Inc., Tokyo). The optical system of the FZP-PC-STEM is shown schematically in Fig. 2. Details of the image calculation are provided in Appendix A. The image of the FZP is formed in front of the specimen position by the condenser lens, and the 0th order diffracted wave converges on the specimen. Around the convergence point on the specimen plane, a weak disk-shaped distribution of the electron beam is also formed by the +1st and −1st order diffracted waves. These interfere with the 0th order diffracted wave, which undergoes a phase change due to the specimen at the center of the probe, whereas the ±1st order waves have an average phase change throughout the disk-shaped probe beam. Therefore, the phase difference between these waves is detected as a contrast in the interference pattern at the rear detection plane. Close to the focal point of the +1st diffracted wave, the 0th- and +1st order diffracted waves dominate the interference pattern near the center. This disk-shaped wave corresponds to the reference wave in electron holography.

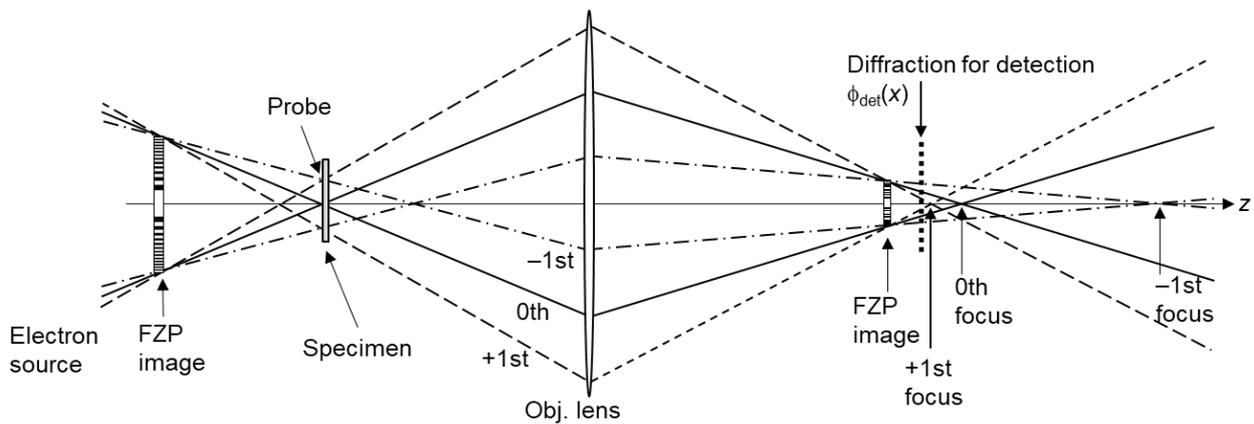

Fig. 2 Schematic diagram of the optical system, showing the relationship between the FZP image, the specimen, and diffraction for detection. Solid-, dashed-, and alternating dot-dash lines indicate the 0th, +1st, and −1st order diffracted waves of the FZP, respectively. The 0th diffracted wave (main beam) converges on the specimen, whereas the ±1st diffracted waves are dispersed in a disc-shaped distribution. After transmission through the specimen, a structured beam is formed on the detection plane by the objective lens. In the system that we used, the diffraction detection plane was set near the focal point of the +1st diffracted wave (thick arrow).

Among the potential detection planes that could function for the FZP-PC-STEM, we selected a defocused plane near the focal point of the +1st order wave for our experiment, to avoid severe beam damages to the high-sensitivity image sensor. To realize this imaging condition, the lens conditions were changed from the default settings for the STEM mode. However, because the diffraction patterns easily drifted on the recording plane during the probe scan, the position and magnification of the diffraction patterns were carefully adjusted with the lenses and deflector of the illumination system to record all the diffraction patterns with the assistance of the shift-correction mechanism of the STEM. Diffraction patterns were collected by using the diffraction imaging technique, though small drifts during the diffraction patterns still remained, even after shift correction. We were also able to acquire a conventional STEM image under these corrected conditions.

In relation to the setup of the FZP-PC-STEM, we first confirmed that the electron beam was focused on the specimen by using a Ronchigram after completing normal beam alignment in the conventional STEM mode. This lens condition was maintained while the area for the FZP-PC-STEM measurement was initially imaged in the conventional STEM mode. Next, after the condenser aperture was switched to the FZP and the objective aperture was removed, a diffraction imaging operation was performed by using a Digiscan unit controlled by Gatan Digital Micrograph Ver. 3 software (GMS; Gatan Inc., Pleasantville, CA). The diffraction images were recorded with a direct electron detector, (K2 summit; Gatan Inc.) in a linear mode. The capture size was 462 x 479 pixels with 4 × 4 binning, and the recording time was 0.1 s per scan point. The FZP-PC-STEM images had 80 × 80 or 100 × 100 scanning points, and the total recording time was 10 or 17 minutes, respectively. The diffraction patterns recorded by the diffraction imaging technique were aligned by using the 'Align SI by peak' function of GMS, and then a mask specially designed for the diffraction pattern was applied by using the GATAN script. The FZP-PC-STEM images were obtained from the masked dataset by using the 'Signal (Dynamic)' function of GMS. The total time for these processing operations was 10–20 minutes.

## 3.   Results and discussion

To investigate the actual probe shape on a specimen, a TEM image was taken with the same lens currents for the illumination system as those for the FZP-PC-STEM described in the Section 2.2. The calculated and experimental images are shown in Figs. 3(a) and 3(b). Details of the calculation procedure are described in Appendix A. In the experimental image, a beam stopper used to protect the sensor from the strong center of the beam can be seen. Fig. 3 show a spiral shape of 12-fold rotational symmetry, corresponding to the FZP pattern shown in Fig. 1, and the features around the center of the experimental image agreed well with those in the calculated image. In contrast, the fine features in the outer part were weak or invisible in the experimental image. In addition, the calculated image suggests that the strong diffuse center spot of the experimental image actually consisted of multiple concentric rings. The vortex feature within ~70 nm of the center was clearly identified in the experiment image.

TEM observation of the FZP showed that the electron transmittance in the Au (30 nm)/SiN (30 nm) film area was ~4%; this formed an incoherent background at the center of the probe. Considering the ratio of the Au/SiN film area (10000 $\mu m^2$) to the area of openings (94 $\mu m^2$) on the FZP, together with the 4% transmittance, the incoherent background in the center of the probe is about four times higher than that of the coherent beam. We speculate that the intensity profile of the experimental image (Fig. 3) does not quantitatively match that of the calculated image because of this incoherent background.

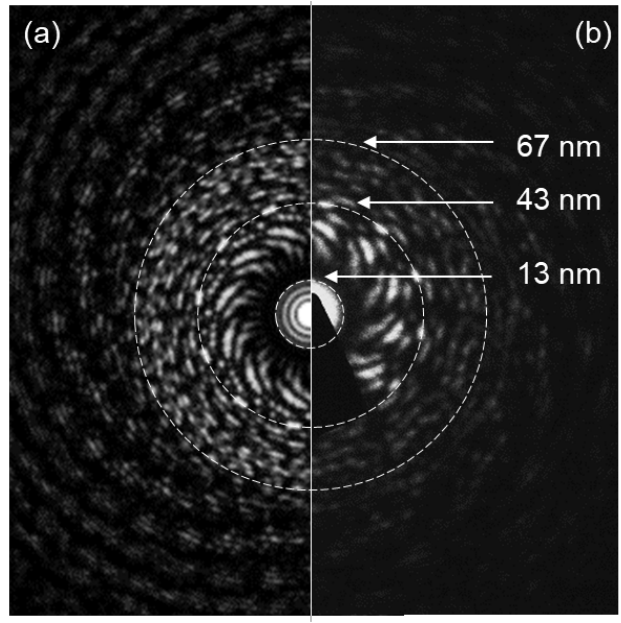

Fig. 3 Intensity distribution of the probe beam at the specimen position: (a) calculated and (b) experimental images. A shadow of the beam stopper can be seen from the center to the bottom in (b). The dashed circles indicate the sizes of the characteristic vortex patterns.

The diffraction patterns at various detection planes were calculated [see Appendix A; Fig. A-3] and compared with those in the experimental images. The position of the detection plane used in this experiment agreed well with that of the defocus amount of $\Delta = -0.195f$, where $f$ is the distance between the FZP image plane and the focal plane of the 0th order diffraction wave. Fig. 4 shows that this calculated image most closely matches the experimental image. The twelve-gear-shape (shown by the black arrow in Fig. 4) of the vortex in the center and the pair of strong and weak spots (white arrows in Fig. 4) in the calculated image can be clearly seen in the two images. Among the six center rings seen in the calculated image, only three were identified in the experimental image. Possible causes of the image degradation in the experimental image are distortions of the FZP during FIB manufacturing, the incoherent background discussed above, and other unconsidered parameters in the calculated image. Further investigation is necessary to clarify this point.

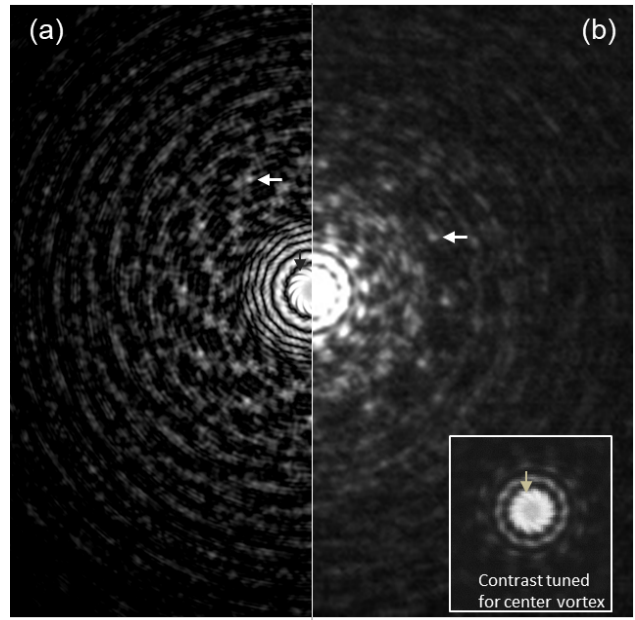

Fig. 4 Diffraction pattern on the detector plane without a specimen: (a) calculated and (b) experimental images. The inset is a contrast-tuned view of the experiment image, clearly showing the shape of the vortex, as indicated by the gray arrow. White arrows indicate characteristic pairs of strong and weak spots in the calculated and experimental images.

Fig. 5 shows results of observations of CNT bundles. The area indicated by the box in the conventional STEM image [Fig. 5(a)] was imaged by the FZP-PC-STEM (80 × 80 scanning points) using the diffraction imaging technique. As a simple post-processing, we first tried to use a single-hole mask [Mask A in Fig. 5(b), top] to extract only the center part of the diffraction patterns in the diffraction imaging dataset. This corresponds to a bright-field image in conventional STEM. The processed image [Fig. 5(c), top] showed strong artificial fringes along the CNTs due to the spread probe shown in Fig. 3. Therefore, by assuming that the CNTs, which consist of carbon atoms, produce a weak electron scattering, we designed a multi-slit mask to reduce the fringing, as described in Appendix B. Fig. 5(c, bottom), is an FZP-PC-STEM image obtained by using the multi-slit mask B [Fig. 5(b, bottom)]. This shows that the image quality is significantly improved and, in contrast to the result with mask A, the strong fringes disappeared around the bundles. The contrast of the CNT intersection area is discussed numerically for these images. Here, the contrast $C$ is defined as $C = (I_{bg} - I)/I_{bg}$, where $I$ is the minimum intensity of the CNT bundle image and $I_{bg}$ is the average image intensity of the thin support film. The values at the points indicated in Fig. 5(a) are summarized in Table 1. The FZP-PC-STEM image formed with mask A showed a high contrast of 63% at point 0. On the other hand, the ZPC-PC-STEM image formed with the multi-slit mask B showed a contrast of 20% at point 0. Note that the contrast value 20% at the intersection, point 0, almost agreed with the sum of the contrasts of two CNT bundles at point 1 (9.3%) and point 2 (13.4%). This result suggests that the image contrast of the FZP-PC-STEM using mask B was proportional to the mass-thickness. A lower-magnification view of the CNTs obtained by using the FZP-PC-STEM can be seen in the SI, which can be found online at http://dx.doi.org/xxxxxx

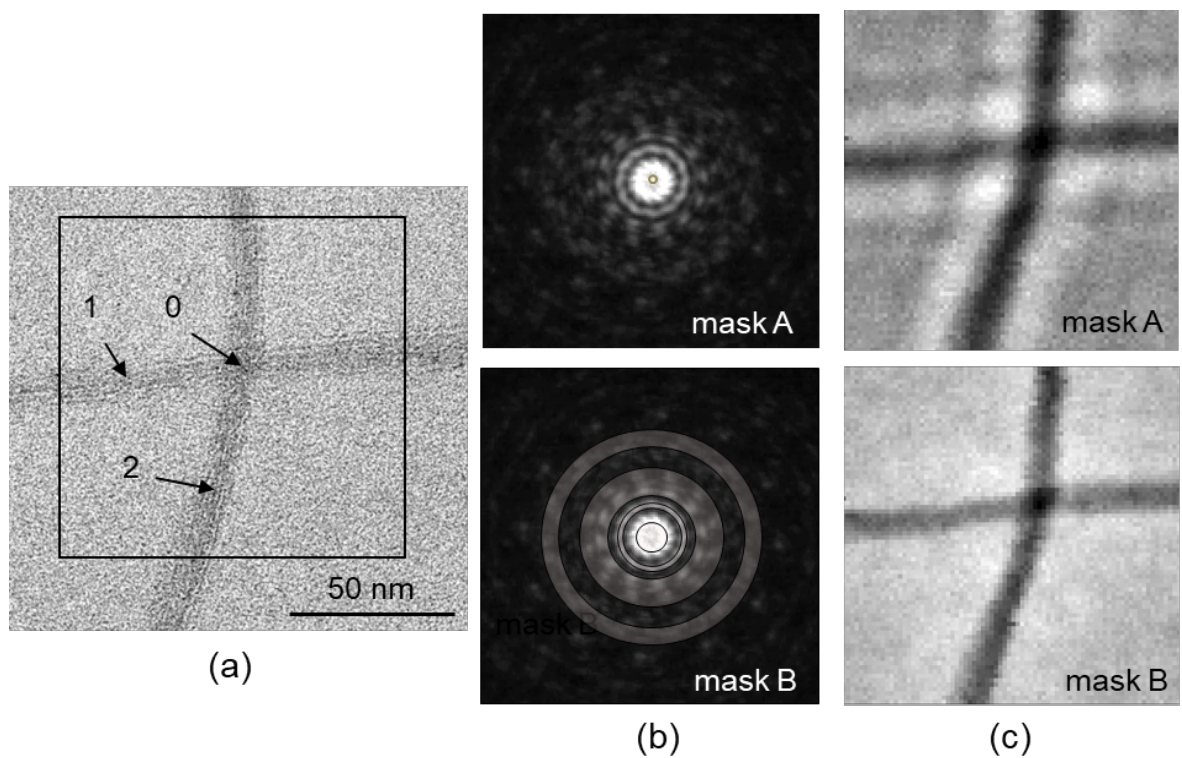

Fig. 5 (a) Conventional STEM image of an intersection area of the CNT bundles on an amorphous thin support film. (b) Mask A having a single hole in the center of the diffraction (top) and mask B consisting of annular slits (bottom). (c) FZP-PC-STEM images of the boxed area in (a) using mask A (top) and mask B (bottom). Although the contrast of the CNT bundles was high with mask A, the contrast obtained by using mask B showed quantitative results without large fringe artifacts.

Table 1 Image contrast values at positions 0, 1, and 2 marked in Fig. 5(a) of the FZP-PC-STEM images using masks A and B.

|          | contrast |        |
|----------|----------|--------|
| position | mask A   | mask B |
| 0        | 62.8%    | 20.2%  |
| 1        | 34.1%    | 9.3%   |
| 2        | 45.0%    | 13.4%  |

Increasing the defocus value of an objective lens enhances low-spatial-frequency components in conventional STEM. The intersection area of the CNT bundles was also imaged by changing the defocus values of the objective lens in the conventional STEM mode (Fig. 6). For the purpose of comparison, all images were normalized so that the average intensity value of the thin support film was 1; that is, each image was divided by the fitted quadratic intensity distribution. In the conventional STEM images, the fringes along the CNT bundles become thicker as the defocus increases. The contrast at the center of the intersection area of the CNT bundles [point 0 in Fig. 6(c)] showed a maximum value (about 12%) when the defocus was $\Delta F = -3000$ nm.

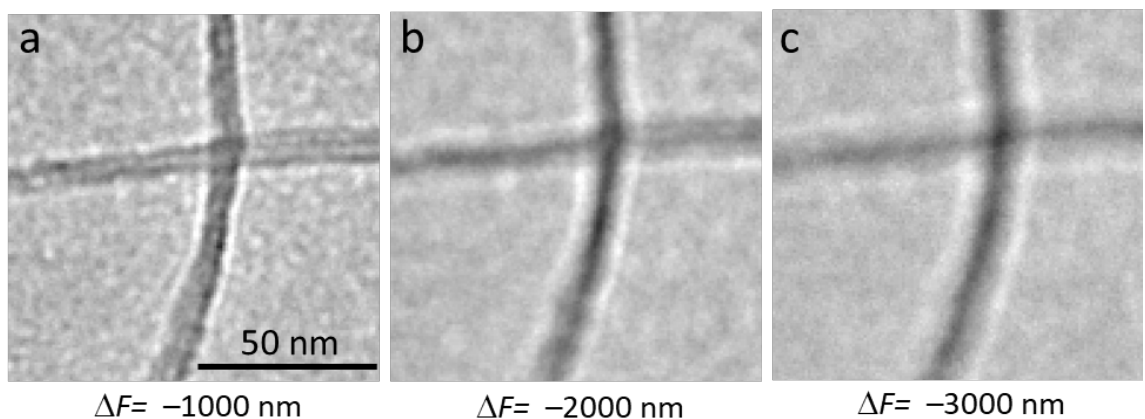

| | | |
|:---:|:---:|:---:|
| a | b | c |
| $\Delta F= -1000$ nm | $\Delta F= -2000$ nm | $\Delta F= -3000$ nm |

Fig. 6 Conventional STEM defocus series for the intersection area of the CNT bundles. The defocus values ($\Delta F$) were (a) −1000 nm, (b) −2000 nm, and (c) −3000 nm.

To compare the spatial frequency components between FZP-PC-STEM and conventional STEM images, the power spectrum of each image was calculated as shown in Fig. 7, where the spatial frequency $k$ is $1/d$ ($d$ = distance in nanometers in the STEM images). At $k > 0.1$ nm$^{-1}$ ($d < 10$ nm), the intensity of the FZP-PC-STEM image was almost the same as or lower than that of the conventional STEM image. However, at $k < 0.1$ nm$^{-1}$, the FZP-PC-STEM image showed a higher intensity. We concluded that the FZP-PC-STEM provided a better image for bundles of CNTs with a width of 10–20 nm, with a quantitative contrast corresponding to the mass-thickness.

Experimentally, the detection plane of the interference pattern shown in Fig. 2 can be set slightly away from the current detection plane or it can be set near the plane where the 0th- or −1st order waves are focused. The interference pattern eventually changes according to the detection plane. For example, the image contrast of the FZP-PC-STEM was inverted when the detection plane was set at a position opposite to the current position with respect to the +1st order convergence plane (r$_{+}$ in Fig. A-1). Because we have not examined other detection planes for the FZP-PC-STEM due to the time required to optimize each set of lens and deflector conditions, computer simulations will be used in efficient determinations of the optimal imaging conditions in the future.

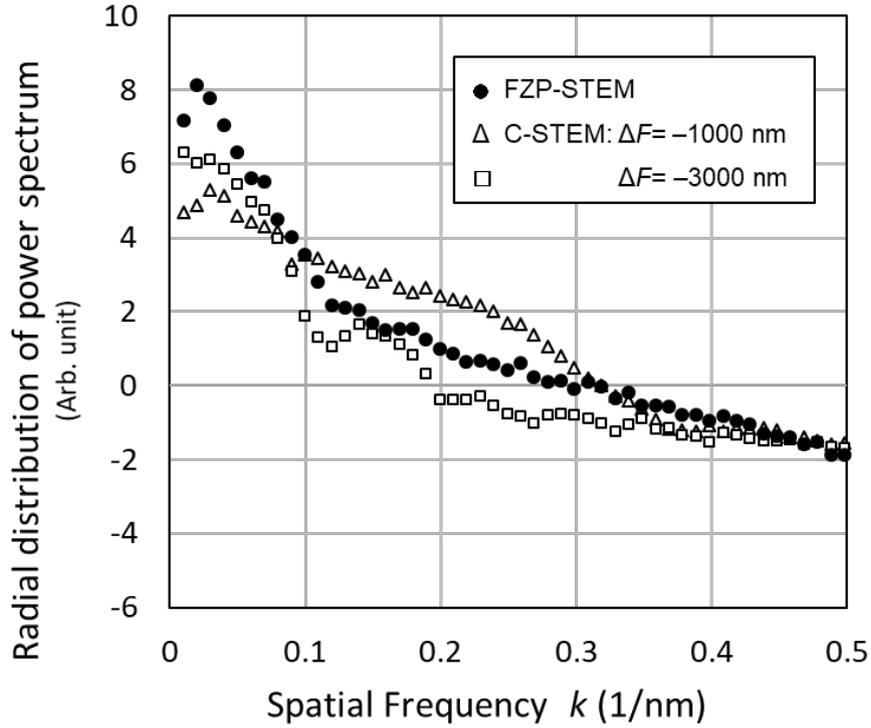

Fig. 7 Radial distribution of the power spectrum for each STEM image (FZP-STEM (●), conventional STEM (C-STEM) at defocus values of ΔF = −1000 nm (△) and −3000 nm (□))

Grillo et al. [9] pointed out that phase modulation of an electron wave by transmission through thin-film materials involves the scattering and absorption of electrons, which inevitably reduces beam quality. VPP is currently widely used, especially in determinations of the structures of many biomolecules. However, the beam transmission through materials and instabilities in the phase shift due to small variations in film thickness limit the achievable resolution of VPP [10]. The situation is more serious with a phase FZP. Even for the latest microfabrication technologies, it is challenging to fabricate perfect phase FZPs, which require a uniform thickness at a sub-nanometer scale. Compared with the phase FZP used in MIDI-STEM [6], an amplitude FZP that consists of simple annular slits can be processed relatively easily. In addition, a gold coating on the amplitude FZP can significantly reduce charging and its associated contamination. In our FZP-PC-STEM, no instability was seen during operation. Furthermore, in our experience, the amplitude FZP introduced in Fig. 1 showed no noticeable deformation or deterioration, even after more than 500 hours of use. We believe that the amplitude FZP used for the FZP-PC-STEM has significant advantages over a phase FZP in terms of the ease of fabrication and operation.

With the current settings for the FZP-PC-STEM, it takes about seven minutes to acquire a dataset of 80 × 80 pixels by using the Gatan K2 camera system, and subsequent data processing takes about ten minutes. Unfortunately, it is impossible to estimate the exact location of data acquisition or the quality of the final image until the process is completed. Increasing the number of scanning pixels to achieve a higher resolution or a wider field of view would require additional acquisition time. Specimen drifts and deformations are expected to be even more serious during such long acquisitions. These issues could be solved if the following two improvements could be achieved in the future: (1) precise positioning of an optimized physical mask in front of the conventional STEM BF detector, and (2) complete fixing of the movement of the diffraction pattern during the beam scan by an improved beam-compensation mechanism.

We believe that these future technological innovations would permit direct observation of FZP-PC-STEM images with the same operations as those used in conventional STEM. Compared with other diffraction imaging-based methods that require post-processing of large datasets, this method has potential advantages in terms of its high-throughput imaging and its ease of operation.

## 4. Conclusion

An amplitude FZP for electrons was manufactured by FIB, and a new phase-contrast observation method (FZP-PC-STEM) using a structured probe beam was proposed and demonstrated. The intensity distribution of the structured probe beam and the detected diffraction pattern agreed qualitatively with the calculated data. By observing CNT bundles with the FZP-PC-STEM using the diffraction imaging technique, it was experimentally shown that the contrast values of low-spatial-frequency components were higher than those in conventional STEM and were proportional to the mass-thickness. In principle, this method does not require a diffraction imaging data-acquisition system, so that it can be introduced into a STEM column as a stable high-contrast imaging method for light-element specimens without major equipment modifications.

## 5. Acknowledgements

This research is supported by the JST ERATO, Momose Quantum Beam Phase Imaging Project (JPMJER1403). Part of the work was supported by the Nanotechnology Platform Project (Nanotechnology Open Facilities in Osaka University) of the Ministry of Education, Culture, Sports, Science and Technology, Japan (F-16-OS-0053, F-17-OS-0022). We gratefully thank Dr. Yoshihiro Arai (Terabase Inc., Aichi, Japan) for his technical support and discussions on STEM optics.

**Appendix A: Simplified optics for diffraction calculation and calculated examples**

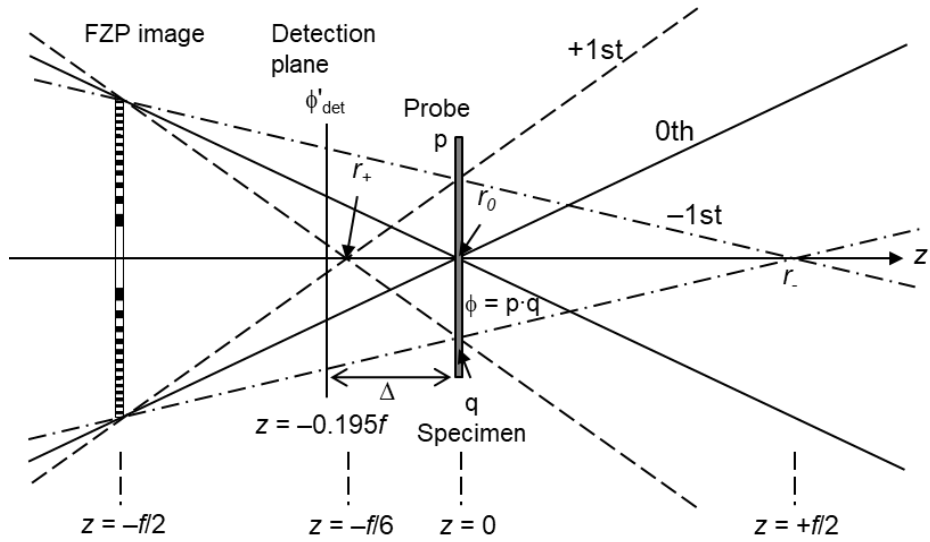

**Fig. A-1** Schematic representation of the relationship between the FZP image, the structured probe function: p, the transmission function of a specimen: q, and the diffraction pattern for detection: $\varphi'_{det}(x)$ in front of the objective lens. The solid line, dashed line, and alternating dot-dash lines indicate the 0th, +1st, and −1st order diffracted waves from the FZP and the corresponding focal points are indicated as $r_0$, $r_+$, and $r_-$, respectively. In the calculations, the wave distribution $\varphi'_{det}(x)$ on the conjugate plane of the detection plane in Fig. 2 was investigated. The distance from the point $r_0$ is expressed as the defocus value $\Delta$ of the detection plane in units of the focal length $f$ of the virtual FZP (FZP image). In the calculations, the distance between the FZP image and the specimen was set to $f/2$, and the $z$ values of some planes are shown at the bottom of the figure ($z = 0$ at the specimen position).

For simplicity, the condenser lens system is omitted from Fig. A-1, and the behavior of each FZP diffraction wave (0th order and ±1st order waves) is shown. As shown in Fig. 2, the wave distribution formed by the objective lens is similar to that in front of the objective lens on the conjugate plane, although the magnification and the convergence angle are different. Therefore, instead of calculating the diffraction wave $\varphi_{det}(x)$ at the detection place behind the lens, we can use the propagation of $\varphi(x)$ at the exit surface of the specimen to the conjugate plane, which is located at the position $z = -\Delta$. This calculation procedure is advantageous in terms of calculation time and reducing errors due to the periodic boundary conditions assumed in propagation calculations using fast Fourier transform.

In the calculation of subsequent diffraction patterns, the same FZP pattern with bridges as that shown in Fig. 1 was generated. The 0th order wave indicated by the solid line in Fig. A-1 is incident on this virtual FZP (FZP image). First, the probe wave function p(x) on the specimen is calculated by propagation from the FZP image ($z = -f/2$) to the specimen position ($z = 0$). To calculate the change in the diffraction pattern induced by the specimen, after multiplying p(x) by the specimen's transmission function q(x), the propagation of $\varphi = $ p·q to the detection plane gives the detected diffraction $\varphi'_{det}(x)$. The propagation function $h$ of Fresnel diffraction was applied in Fourier space. For the spatial frequency $\boldsymbol{k}$ ($kx$, $ky$), $k=|\boldsymbol{k}| = 1/d$, $h$ is given as $h(\boldsymbol{k})=exp(\pi i\lambda\Delta zk^2)$, where $\lambda$ is the electron wavelength and $\Delta z$ is the propagation distance [11]. The programing language that we used was Python (ver. 3.7) with the numpy and cupy packages. The image size was 4096 × 4096. Spatial and temporal coherence and incoherent background were not considered in these calculations.

Fig. A-2 shows the calculated result for the defocus series of diffraction patterns without a specimen. As defined in Fig. A-1, $\Delta z = -0.50f$ is the plane of the FZP image, and $\Delta z = 0$ is the plane of the 0th order diffraction convergence. As indicated by white marks in the figure, it can be seen that the 0th- and +1st order diffracted waves distributed in a disk-shape converge at $\Delta z = 0$ and $-1/6f$ ($\approx -0.167f$), respectively.

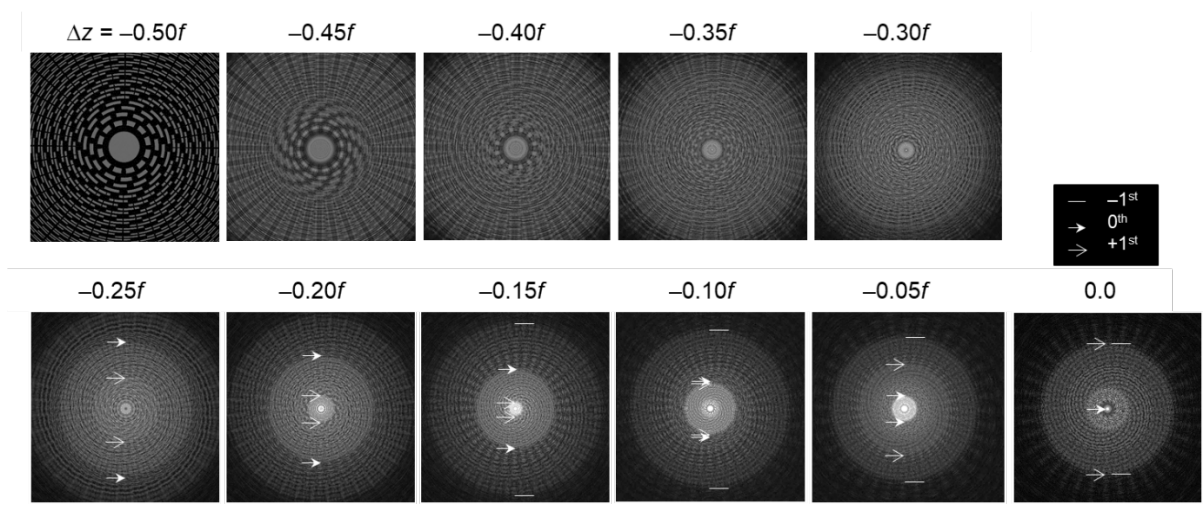

**Fig. A-2** Calculated defocus series of diffraction patterns. $\Delta z = -0.50f$ is in the plane of the FZP image, and $\Delta z = 0$ is in the plane of the 0th order diffraction convergence. The white arrows and lines indicate the convergence and divergence behavior of the 0th and ±1st order diffracted waves.

Fig. A-3 shows the changes in the diffraction pattern in steps of $0.005f$ around $\Delta z = -0.195f$, which was estimated as the position of the diffraction pattern detected in the experiment (Fig. 4). At $\Delta z = -0.205f$, the direction of the central vortex shape was counterclockwise, but it changed to clockwise at $\Delta z = -0.195f$. The center of the vortex component then weakened at $\Delta z = -0.185f$. Because the diffraction pattern was sensitive to $\Delta z$, a wider range of patterns was visually compared with the experimental pattern (Fig. 4) to determine the experimental condition as $\Delta z = -0.195f$. As described in the text, there was a strong incoherent background that converged like the 0th order coherent wave, so that a quantitative comparison cannot be made at present.

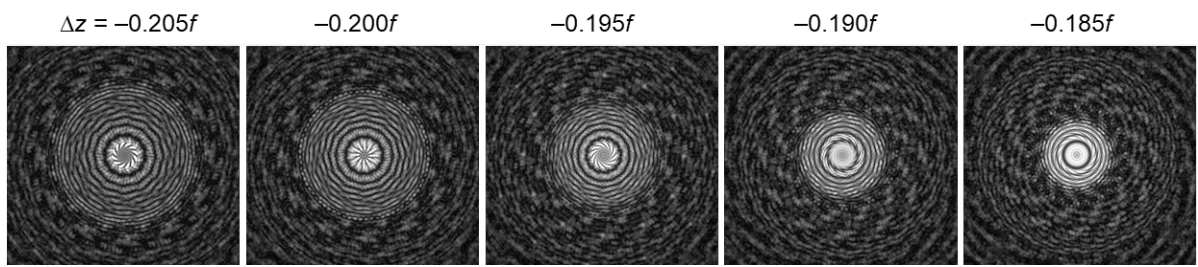

| $\Delta z = -0.205f$ | $-0.200f$ | $-0.195f$ | $-0.190f$ | $-0.185f$ |

**Fig. A-3** Diffraction pattern calculated in $0.005f$ steps around $\Delta z = -0.195f$. Note the sensitivity of the center pattern to $\Delta z$. The spiral structure at the center changes significantly with $\Delta z$ in this range.

## Appendix B: Setting of an imaging mask for FZP-PC-STEM

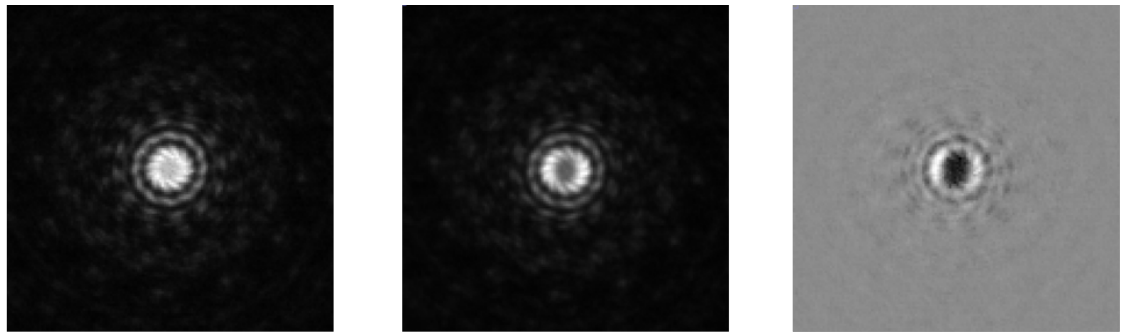

(a) diffraction from a film area    (b) diffraction from CNT bundle    (c) Subtraction (b)-(a)

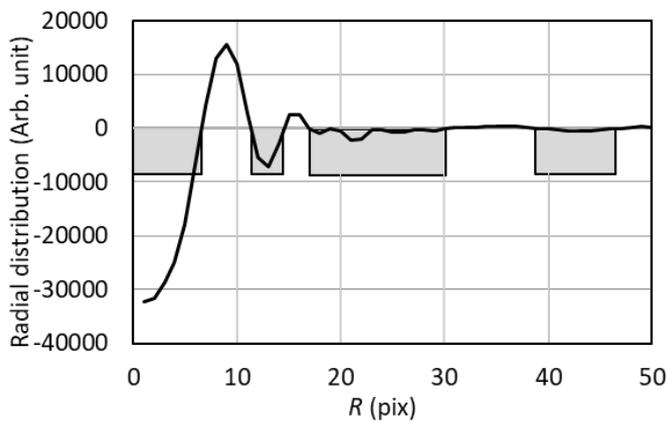

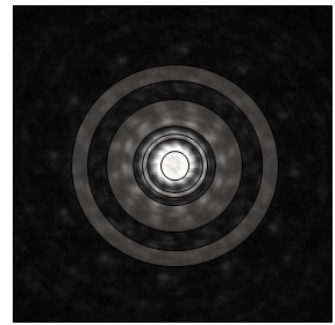

(d) Radial distribution of subtracted diffraction pattern [Fig. (c)]    (e) Determined multi-slit mask
and areas (halftone dot meshed) for 4D-STEM imaging    for 4D-FZP-STEM

**Fig. B** Diffraction patterns of the Fig. 5. Diffraction patterns from (a) an amorphous thin film, (b) CNT bundles on the amorphous thin film, and (c) intensity subtraction (b) – (a). (d) Radial distribution of the subtracted diffraction pattern of (c). Areas with negative values, indicated by a halftone mesh, are used for the FZP-PC-STEM imaging. (e) Multi-slit mask B (halftone dot meshed) designed to correspond with the regions of negative values in (d).

In MIDI-STEM, the diffraction intensity in the detection system corresponding to the shape of the phase FZP is known in advance, and a matched pattern based on this configuration is therefore used. However, our detection plane of the diffraction pattern at $\Delta z = -0.195f$ was not a well-defined plane as in Fraunhofer diffraction. Hence, a multi-annular-slit mask was defined experimentally as follows.

Because a CNT bundle consists of a light element and is thin (~10 nm in diameter), scattering of the electron beam can be assumed to be proportional to the thickness of the CNT bundle on the basis of the weak-phase-object approximation. Furthermore, according to the discussion in Appendix A, the interference between the 0th- and +1st order waves can be assumed to be dominant at the detection plane ($\Delta z = -0.195f$), where the −1st order wave is widely distributed on the detection plane with a radius that is about twice as large. Therefore, the difference between the diffraction patterns from the thin-film portion [Fig. B(a)] and from the intersection of the CNT bundles [Fig. B(b)] was calculated, and the difference due to the CNT bundles was extracted. On the basis of the radial distribution [Fig. B(d)], we designed a multi-annular-slit mask, as shown in Fig. B(e), that masks out the areas of the positive values in the radial distribution for the FZP-PC-STEM imaging. In this report, a script for the annular/disk mask prepared by GMS was used in the processing. In the future, the introduction of a weighting mask based on the peak intensity and symmetry might result in higher-resolution images.

## References

[1]    R. Danev, B. Buijsse, M. Khoshouei, J.M. Plitzko, W. Baumeister, Volta potential phase plate for in-focus phase contrast transmission electron microscopy, Proc. Natl. Acad. Sci. U. S. A. 111 (2014) 15635–15640. https://doi.org/10.1073/pnas.1418377111.

[2]    R. Danev, W. Baumeister, Expanding the boundaries of cryo-EM with phase plates, Curr. Opin. Struct. Biol. 46 (2017) 87–94. https://doi.org/10.1016/j.sbi.2017.06.006.

[3]    R. Danev, K. Nagayama, Transmission electron microscopy with Zernike phase plate, Ultramicroscopy. 88 (2001) 243–252. https://doi.org/10.1016/S0304-3991(01)00088-2.

[4]    W.D. Rau, P. Schwander, F.H. Baumann, W. Höppner, A. Ourmazd, Two-dimensional mapping of the electrostatic potential in transistors by electron holography, Phys. Rev. Lett. 82 (1999) 2614–2617. https://doi.org/10.1103/PhysRevLett.82.2614.

[5]    H. Minoda, T. Tamai, H. Iijima, F. Hosokawa, Y. Kondo, Phase-contrast scanning transmission electron microscopy, Microscopy. 64 (2015) 181–187. https://doi.org/10.1093/jmicro/dfv011.

[6]    C. Ophus, J. Ciston, J. Pierce, T.R. Harvey, J. Chess, B.J. McMorran, C. Czarnik, H.H. Rose, P. Ercius, Efficient linear phase contrast in scanning transmission electron microscopy with matched illumination and detector interferometry, Nat. Commun. 7 (2016) 1–7. https://doi.org/10.1038/ncomms10719.

[7]    H. Yang, P. Ercius, P.D. Nellist, C. Ophus, Enhanced phase contrast transfer using ptychography combined with a pre-specimen phase plate in a scanning transmission electron microscope, Ultramicroscopy. 171 (2016) 117–125. https://doi.org/10.1016/j.ultramic.2016.09.002.

[8]    D. Winston, V.R. Manfrinato, S.M. Nicaise, L.L. Cheong, H. Duan, D. Ferranti, J. Marshman, S. McVey, L. Stern, J. Notte, K.K. Berggren, Neon ion beam lithography (NIBL), Nano Lett. 11 (2011) 4343–4347. https://doi.org/10.1021/nl202447n.

[9]     V. Grillo, J. Harris, G.C. Gazzadi, R. Balboni, E. Mafakheri, M.R. Dennis, S. Frabboni, R.W. Boyd, E. Karimi, Generation and application of bessel beams in electron microscopy, Ultramicroscopy. 166 (2016) 48–60. https://doi.org/10.1016/j.ultramic.2016.03.009.

[10]    R. Danev, H. Yanagisawa, M. Kikkawa, Cryo-Electron Microscopy Methodology: Current Aspects and Future Directions, Trends Biochem. Sci. 44 (2019) 837–848. https://doi.org/10.1016/j.tibs.2019.04.008.

[11]    J. M. Cowley, Diffraction physics, Elsevier Science B.V, 1995.

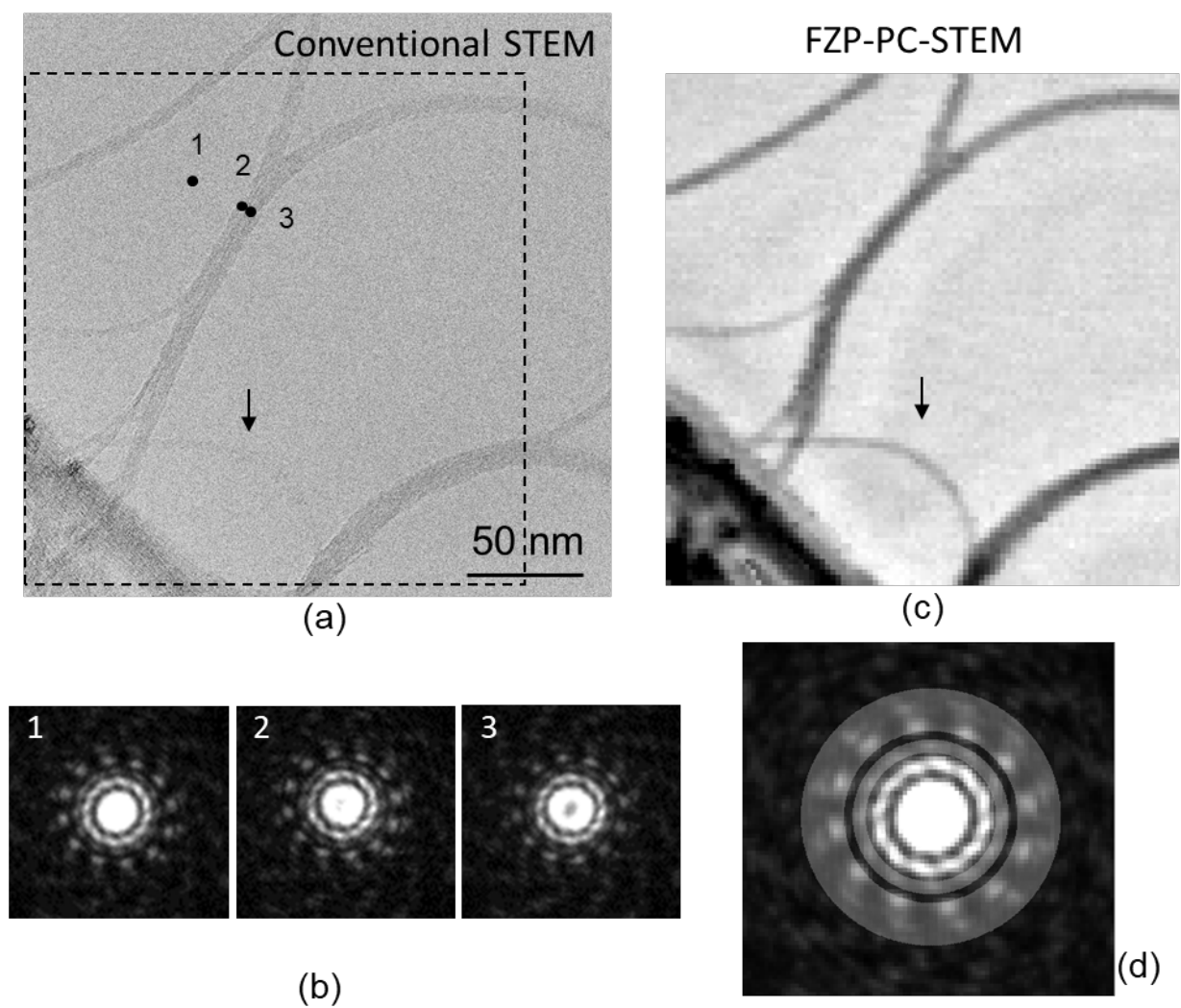

**Fig. S-1** (a) Conventional STEM image of CNT bundles, (b) Detected diffraction patterns from points 1–3 in (a), (c) FZP-PC-STEM image for the boxed area in (a). (d) Detected diffraction pattern with a multi-slit mask (halftone dot meshed) for image processing of the FZP-PC-STEM.

Fig. S-1(a) shows a conventional STEM image of the CNT bundles with a wide field of view. The detected diffraction patterns obtained from points 1–3 in Fig. S-1(a) are shown in Fig. S-1(b). Compared with point 1 with no specimen, points 2 and 3, where the probe center hits the CNT bundle, show spots with darker centers due to a phase shift of the electrons at the probe center. The square area in the conventional STEM was imaged by FZP-PC-STEM using the diffraction imaging technique [Fig. S-1(c)]. The FZP-PC-STEM image (100 × 100 pixels) was formed with the multi-slit mask shown in Fig. S-1(d), which was determined by the same procedure as described in Appendix B. The thick CNT bundle in the lower left-hand corner in the conventional STEM image is significantly larger than the probe diameter, where the phase shift of electrons in the CNT bundle is enormous. Therefore, the contrast in this area is no longer proportional to the actual thickness of the specimen. The thin CNT bundle indicated by the arrow in Fig. S-1(a) had a sufficiently dark contrast in Fig. S-1(c). Although the fine structures were not resolved in the FZP-PC-STEM

image with the current instrumentation setup, the thicker CNT bundles had an apparently darker contrast than thinner ones. Consequently, FZP-PC-STEM images seems to be more sensitive to differences in material quantities than a conventional STEM image.